\documentclass[conference]{IEEEtran}
\IEEEoverridecommandlockouts
\usepackage{cite}
\usepackage{upgreek}
\usepackage{amsmath,amssymb,amsfonts}
\usepackage{algorithmic}
\usepackage{algorithm2e}
\usepackage{graphicx}
\usepackage{textcomp}
\usepackage{float}
\usepackage{url}
\usepackage{subfig}
\usepackage{authblk}
\usepackage{xcolor}
\usepackage{float}
\usepackage{xspace}
\usepackage{tabularx}
\newcommand{\DCSim}{HolDCSim\xspace}
\def\BibTeX{{\rm B\kern-.05em{\sc i\kern-.025em b}\kern-.08em
    T\kern-.1667em\lower.7ex\hbox{E}\kern-.125emX}}

\begin{document}

\title{\LARGE{\DCSim: A Holistic Simulator for Data Centers}}

\author{Fan Yao$^{\dagger}$}
\author{Kathy Ngyugen$^{\ddagger}$}
\author{Sai Santosh Dayapule$^\ddagger$}
\author{Jingxin Wu$^\ddagger$}
\author{Bingqian Lu$^\S$}
\author{\\Suresh Subramaniam$^\ddagger$}
\author{Guru Venkataramani$^\ddagger$}
\affil{$^\ddagger$Department of ECE, The George Washington University\vspace{-4mm}}
\affil{$^\dagger$Department of ECE, University of Central Florida\vspace{-4mm}}
\affil{$^\S$Department of CSE, University of California Riverside}


\maketitle
\pdfminorversion=7
\begin{abstract}

Cloud computing based systems, that span data centers, are commonly deployed to offer high performance for user service requests. As data centers continue to expand, computer architects and system designers are facing many challenges on how to balance resource utilization efficiency, server and network performance, energy consumption and quality-of-service (QoS) demands from the users. 

To develop effective data center management policies, it becomes essential to have an in-depth understanding and synergistic control of the various sub-components inside large scale computing systems, that include both computation and communication resources.
Prior studies on performance and energy issues in data centers largely focus on either servers or the network and completely ignore issues relating to the other components, or consider only high level analytical models without sufficient detail, which can lead to non-optimal solutions.
Unfortunately, it is prohibitively expensive or in some cases even impossible to have complete access to an operational large-scale computing system (e.g., production server farms). Therefore, 
a comprehensive \emph{simulation infrastructure} that models all major hardware and system components, and offers interfaces to manage the interplay between both computation and communication resources
are critical in advancing future research for more effective performance and energy optimization in data centers. 


In this paper, we propose \emph{\DCSim}, a light-weight, holistic, extensible, event-driven data center simulation platform that effectively models both server and network architectures. \DCSim can be used in a variety of data center system studies including job/task scheduling, resource provisioning, global and local server farm power management, and network and server performance analysis. We demonstrate the design of our simulation infrastructure, and illustrate the usefulness of our framework with several case studies that analyze server/network performance and energy efficiency. We also perform validation on real machines to verify our simulator.


\end{abstract}

\section{Introduction}

 Rapid advancements in information technology and growth in user needs over the past decade have significantly pushed for the adoption of large scale server systems. Today's Internet Services typically require a tremendous amount of computation, communication and storage resources. As a result, service providers continue to scale out and scale up the hardware infrastructure. While considerable amount of hardware resources offered in large-scale computing systems continue to enable various application paradigms, they also pose unprecedented challenges when it comes to effectively managing these resources to achieve higher performance, resource utilization and power efficiency. Meanwhile, many of today's cloud services come with service level agreements that specify the expected quality-of-service needs from customers. Optimizing the various aspects in data center systems with the consideration of user-level service agreement is a critical and challenging mission~\cite{barroso_isca07,kanev2014tradeoffs,LiuSleepScaleRuntimejoint2014,lo2014towards,yao2014comparative}. 

While performance and efficiency on single server platforms have been extensively studied by prior work~\cite{MeisnerPowerNapeliminatingserver2009,HoffmannDynamicknobsresponsive2011}, managing scale-out services that leverage task-level parallelism requires understanding of the interactions among major data center components including both server and network resources.  As such, the data center infrastructure typically involves thousands of servers, hundreds of network devices (e.g., switches, routers) and complex network topology configurations. Often times, physical access to these testbeds are very limited, constricting researchers' ability to understand the landscape for optimizing these systems. Thus, a comprehensive simulation platform that models both computation and communication hardware is a necessary tool for effective and rapid modeling of these large scale system configurations, and would allow for \emph{end-to-end, holistic} large-scale data center workload studies including workload characterization, performance engineering and energy optimization. 

Existing data center simulation tools by and large fall into two classes: 1) simulators that model details of the machine hardware (e.g., gem5~\cite{BinkertGem5Simulator2011}, zsim~\cite{SanchezZSimFastAccurate2013} and dist-gem5~\cite{dist-gem5}). These tools are able to provide cycle-accurate simulation with fine-grained performance and power models; 2) tools that simulate a cluster of servers in a distributed setting where each individual server is modeled as a pool of resource slots for task scheduling~\cite{CalheirosCloudSimNovelFramework2009,TigheDCSimdatacentre2012, bighouse}. These frameworks allow modeling of high-level resource management with the goal of improving server-level or network-level performance and energy efficiency. While these simulators have been demonstrated to be useful in several use cases, they have potential shortcomings that make them {\it less effective} for {\it comprehensive} studies of data center systems: (1) While architecture simulators give high-resolution statistics on many microarchitectural level components, they typically take hours or days to simulate even one-second execution in real systems. This makes them unsuitable to capture macro-level operations for distributed applications running in data center systems. (2) Existing cloud simulators either model server-only~\cite{bighouse} or network-only data center components~\cite{ns2}, or they model only at a very high level largely abstracting away hardware characteristics (e.g., low power state for line cards)~\cite{CalheirosCloudSimNovelFramework2009,lim2009mdcsim} that make them less effective in simulating fine-grained hardware activities for certain applications such as latency-critical workloads. 

As data center systems continue to evolve, it is crucial to understand application performance and power characteristics by considering both server and networking hardware, along with their interactions as a whole. 
The design of a simulation platform should incorporate two aspects to close the existing gaps:
\begin {itemize}
\item There needs to be an appropriate level of abstraction for servers that sufficiently exposes  hardware-level knobs (e.g., core- and package-level power management features) that could be leveraged by high-level system components. This will close the gap between architectural simulation platform \emph{where many hardware low-level features are simulated but may be unnecessary for study of large scale systems} (e.g., statistics such as branch prediction accuracy, mis-speculation flushes in the pipeline), and current cloud computing simulators \emph{where only high-level tasks and hardware resource capabilities (e.g., VMs and number of cores) are modeled but are lacking mechanisms to realistically control or study these features} (e.g., incorporate sleep states to manage power vs. performance tradeoffs, ability to do performance analysis with realistic network traffic flow models between jobs running in a data center system).

\item The simulator will need to jointly incorporate computation and communication in a systematic manner in order to enable full control of the system across the hardware and software stack. This essentially bridges the gap between cloud and network simulation platforms by integrating important communication aspects into cloud applications for more realistic modeling of such systems.
\end{itemize}

To satisfy the above two design goals, there is an {\it acute} need to build a holistic data center simulation infrastructure that completely models all of the critical components including servers and their processors, network switches and configurations (topologies, workloads, traffic patterns). 

In this paper, we present \emph{\DCSim}, a lightweight, highly-extensible event-driven data center simulator that jointly simulates data center servers and networks.  \emph{\DCSim} efficiently models the components of server and network devices with necessary hardware details (queuing, dynamic power management and idle power management) to sufficiently capture end-to-end application performance and power characteristics. We systematically demonstrate the various modules in \emph{\DCSim} along with our design considerations. \DCSim considers queuing effects from both network and servers, and accounts for performance based on various sources of delays in them. Finally, it provides interfaces to access many state-of-the-art power management features that are available in modern server and network switch systems, which enable future studies on joint server and network power optimizations.

We demonstrate the use of \DCSim with four detailed case studies: (1) data center level system resource provisioning, (2) managing active servers by monitoring utilization levels, (3) fine-grain adaptive power optimization policy using processor low power states to improve energy efficiency for latency-critical workloads with QoS constraints, and 4) joint server-network low power state management that judiciously coordinates server and network resources.

\begin{table*}[htb]
	\centering
	\small
	\begin{tabular}{|c|c|c|c|} 
		\hline
		& \textbf{\DCSim} & \textbf{BigHouse~\cite{bighouse}} & \textbf{CloudSim~\cite{GargNetworkCloudSimModellingParallel2011}} \\
		\hline
		Server& Multi-core processor; Multiple sockets & Multi-core processor; Multiple sockets& Multi-core processor\\
				 &  Supports Heterogeneous architecture& & \\
		\hline
		Network& Models switch, linecards and ports&No switch model& Models switch, link and ports\\
		\hline
				      & Switch-only (e.g., fat tree~\cite{al2008scalable})&  & Use BRITE topology generator\\
		Topology & Server-only (e.g., CamCube~\cite{camcube})& No topology model  &  \\
			          & Hybrid-only (e.g., BCube~\cite{bcube})& & \\
		\hline
		Communication & Packet-level & No communication model&  Packet-level\\
		                         &Flow-based & & \\
		\hline
		Job/Task& Multi-task job& Single-task job&  Multi-task job\\
				     & Supports task-dependency DAG& & \\
		\hline
		& Per-core DVFS& &  \\
			     & Core and package sleep states& & Extension models\\
		Power	     & ACPI system sleep states&Single processor low power state  & server system sleep states~\cite{XavierModelingsimulationglobal2017}\\
			     & Switch link rate adaption& & \\
			     & Switch port and linecard low power states& & \\
		\hline
		Scalability & More than 20K servers & $<$1K servers & About 1.5K servers\\
		\hline
	\end{tabular}
	\caption{Comparison of \DCSim with two widely-used simulators.}
	\label{tab:table1}
	\vspace{-5mm}
\end{table*}
\section{Motivation}

\begin{figure*}[ht]
	\captionsetup{font=small}
	\centering\includegraphics[width=0.8\textwidth]{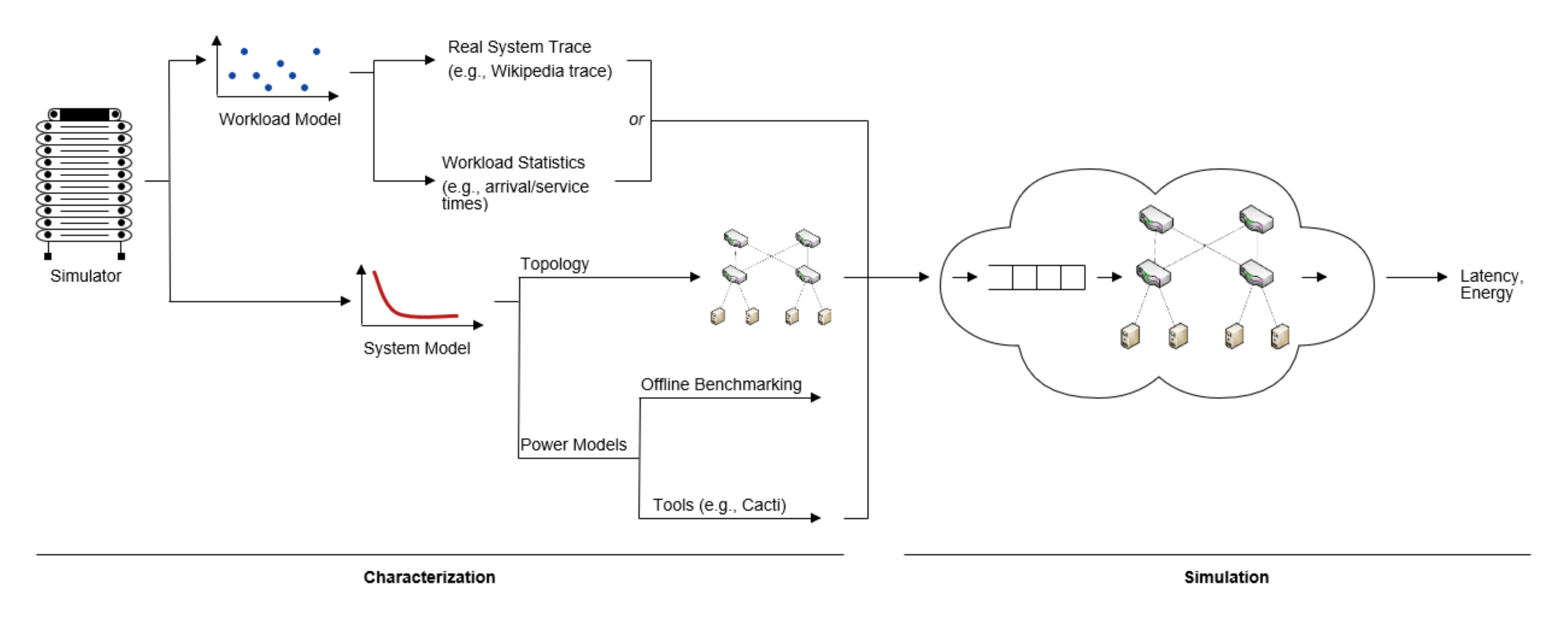}
	\caption{Overview of the design of \DCSim}
	\label{figure1}
	\vspace{-5mm}
\end{figure*}

To motivate the need for holistic modeling of servers and network in data centers, we will consider the issue of improving the job response time and energy efficiency for latency-critical workloads in data centers, which have been studied by several prior works~\cite{KastureRubikFastanalytical2015a,LiuSleepScaleRuntimejoint2014,GandhiAutoScaleDynamicRobust2012,YaoDatacenterspower2012}. 
Specifically, when a job request is received, the job is interpreted as a set of tasks to be executed. The data center \emph{front-end} will schedule the tasks onto one or several servers. The task request then traverses through network links and switching devices until it arrives at the \emph{back-end} server. This procedure may encounter congestion on links, ports and switching fabric (e.g., line cards).
Upon receiving the request (i.e., network packets), the server hardware (e.g., processors and DRAM) may need to be woken up if they have entered low-power states. Once activated, servers will allocate their available resources (e.g., cores, memories and I/O) for the task to execute. The results of the task are forwarded to its dependent tasks and eventually the job completes with final results sent back to users. 

Based on the above description, we know that the latency for a job has two major sources -- network latency and server latency. Server latency can be further divided into hardware latency, queuing latency and computation latency. 
As servers continue to offer higher performance and power efficiency, network latency and power consumption become a significant component~\cite{Abts:2010:EPD:1815961.1816004}. This is especially true for latency-critical workloads that form many critical distributed services. 

Motivated by the aforementioned observations, we argue that in order to formulate effective system power policies and performance tuning, it is important for the simulation platform to provide the following capabilities:
\begin{itemize}
	\item \textbf{Modeling of server processors with the ability to include hardware-level parallelism.} Such modeling is important as multi-core processors are ubiquitous in modern computing systems, especially in the cloud. Multi-core applications exhibit new challenges as compared to traditional single-thread processing due to the effects of interference and resource sharing. Additionally, heterogeneous processors with performance varying cores should also be considered due to their advantages in bringing better performance-power tradeoff.
	\item \textbf{Modeling of global and local job scheduler.} Today's data center systems commonly feature a global-level job scheduling module that is responsible for dispatching users' job requests to individual servers. Moreover, there is typically a local scheduler within each server which manages task dispatching to each execution unit.  Local scheduler is application dependent, and several prior works have shown the performance impact of local scheduler policies (e.g,  a unified task queue or per-core task queue~\cite{LiTalesTailHardware2014}).
	\item \textbf{Inclusion of power mode control mechanisms.} Particularly, the simulator should provide commercial off-the-shelf power management features that can be leveraged for enhancing idle power and dynamic power consumption. The power model should be constructed in a hierarchical fashion that takes into account the core, package and socket infrastructure for servers and power, line card, and switching fabric for network switches. 
	\item \textbf{Integrated modeling of servers, network devices and topology.} Existing research has shown that packet switching and network topologies play a crucial role in determining the overall application performance, and hence the ability to model their performance and power consumption in conjunction with servers is a key consideration to accurately understand overall performance of data centers. 
\end{itemize}

We aim to equip \DCSim with all the above-mentioned capabilities. Table~\ref{tab:table1} compares \DCSim with two state-of-the-art simulators, one that models servers in detail (BigHouse) and another that models distributed cloud systems (CloudSim).

\section{\DCSim System Design}
At a high level, \DCSim has three major components: \emph{workload module}, \emph{server module}, and \emph{network topology/switch module}. The \emph{workload generator module} generates load to the simulated data center by injecting job requests. Each job represents a user's service request (e.g., a search query~\cite{barroso2003web} or a request for cached content~\cite{memcached}) that will lead to a sequence of executable tasks performed in the back-end servers. 
The server module instantiates a cluster of servers based on a configurable user script. Each server has one or several multi-core processors, a DRAM component and other platform resources (e.g., disks and power supply unit). Servers schedule local hardware resources including core and memory to each task. Each core is considered as a processing unit that can serve one task at a time. Core performance is determined by its hardware configuration (e.g., operating frequency) and task settings (i.e., computation intensiveness). Additionally, we build an extensive ACPI-based power model for servers, and provide for a fine-grained power management for processors and their peripheral hardware. 

\DCSim models a complete data center infrastructure by modeling network devices (e.g., switches) and interconnection among various nodes in the system (i.e., topologies). The network module creates a complete topology by connecting the switches and servers with network links. Network communication is modeled at two levels of granularity: packet-based communication and flow-based communication. 
Each network switch contains several key hardware components including \emph{ports}, \emph{line cards} and \emph{chassis}. \DCSim models packet queuing and packet-forwarding for each switch. Similar to servers, we also build a power model considering various switch power management features such as Low Power Idle (LPI~\cite{2005LuoLowpowernetwork}) and dynamic link rate adaptation~\cite{2008GunaratneReducingEnergyConsumption}.


Figure~\ref{figure1} illustrates the high-level workflow design for  \DCSim. \DCSim takes a workload model, server and switch profile as inputs to run experiments. During simulation, \DCSim keeps track of several types of runtime statistics such as power and energy consumption, network delays, job latency, and power state transitions. 






\subsection{Server Architecture}
\begin{figure}
	\centering\includegraphics[width=0.3\textwidth]{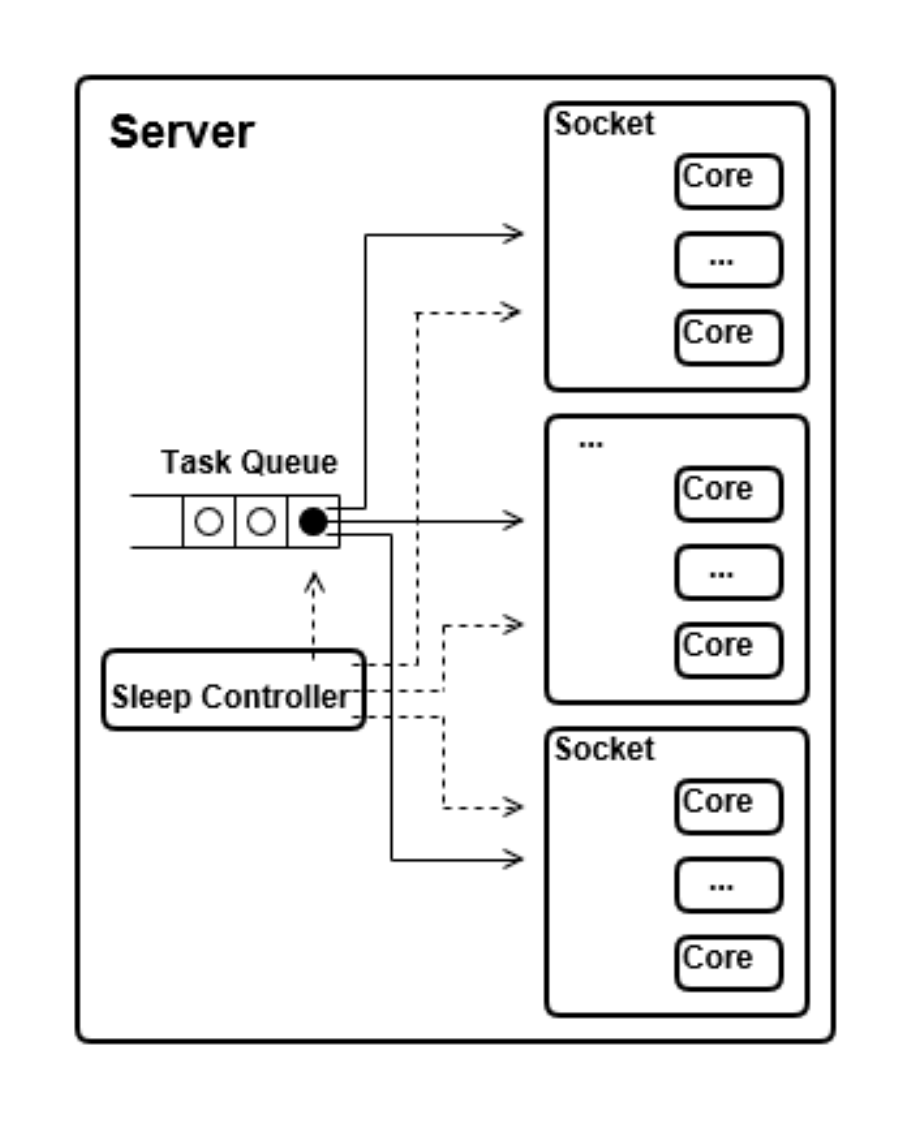}
	\caption{Overview of server model in \DCSim.}
	\label{figure2}
\end{figure}

 Cores have the capability to enter multiple levels of power states. Each server maintains a local queue where all incoming task requests are buffered. Meanwhile, each core can also have its local task queue, and serves one task at a time. Queuing delays are taken into account for task processing latencies. The task processing time for each task is determined by the service time of the task and the operating frequency of the core on which it executes. Note that we also model various types of workloads with different levels of computation intensiveness. 

\DCSim models server power based on the widely deployed Advanced Configuration and Power Interface (ACPI standard) ~\cite{acpi_linux}. ACPI uses global states, ${G_x}$, to represent states of the entire system. For each ${G_x}$ state, there is one or more system sleep states, denoted as ${S_x}$. System sleep states define power status for various server components. When the system sleep state is ${S_0}$, the processor is allowed to reside in a set of C states such as C0. C state enables fine-grained core and uncore components (shared caches and coherence fabric) low-power modes to achieve various levels of power savings. Finally, performance states can be configured to determine the speed of instruction execution at runtime (i.e., DVFS). Modern processors generally provide high parallelism by integrating multiple cores within a processor package. Low-power C states are supported at both core- and package levels that are modeled in \DCSim. 


\subsection{Switch and Network Architecture}
\begin{figure}
	\captionsetup{font=small}
	\centering\includegraphics[width=0.3\textwidth]{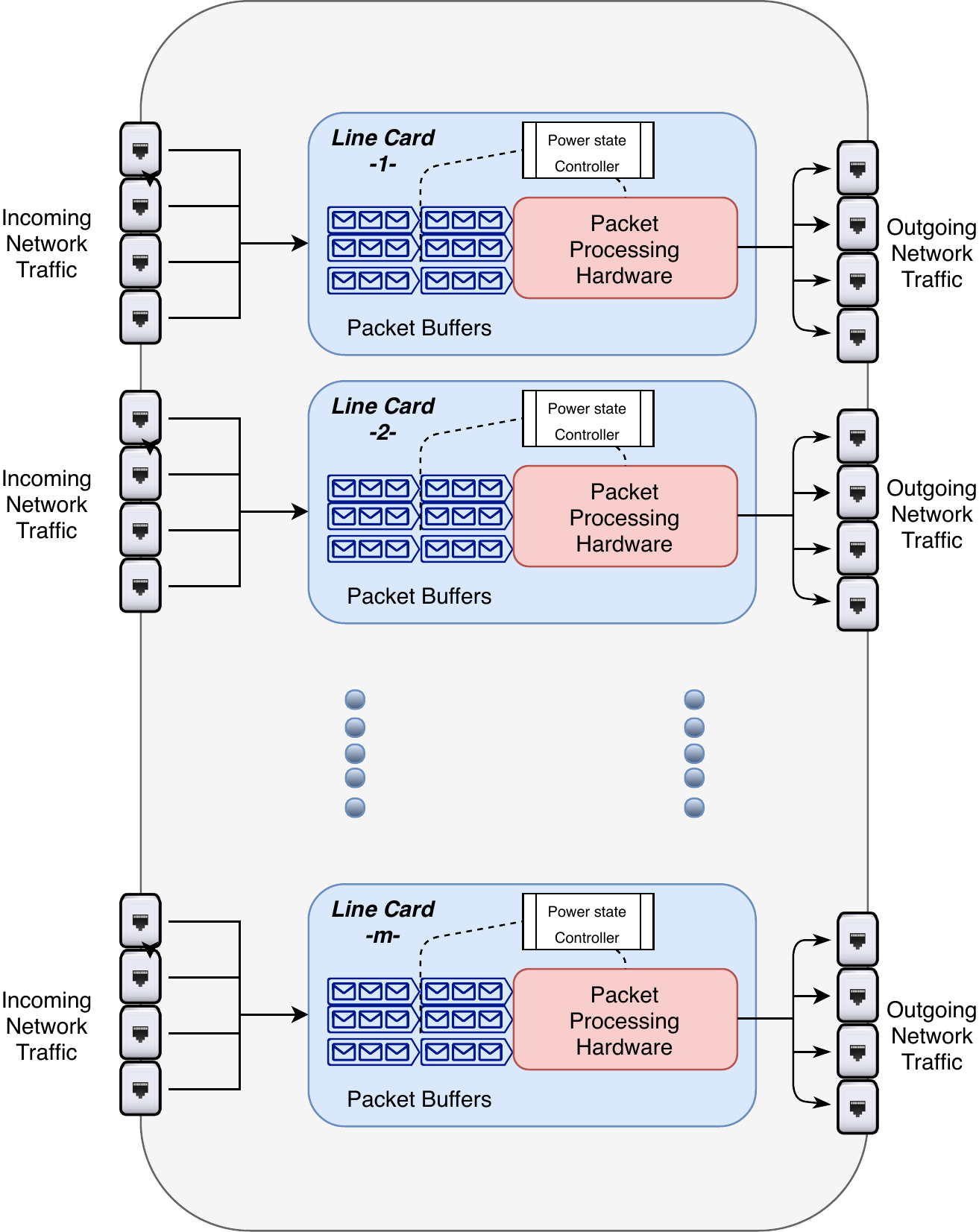}
	\caption{Overview of switch model in \DCSim.}
	\label{figure3}
\end{figure}

The profile of network switches is modeled from real systems and online power modeling tools. Network switches can have multiple line cards, and line cards can have many ports. Configurable parameters of line cards include number of ports and power consumption in different power states. The link rate, buffer size, power state, and transition delay can be configured for each port. 


In our simulator, we assume three power states for each switch port: active, LPI (Low Power Idle~\cite{christensen2010ieee}), and off state. Currently, we are also assuming three power states for each switch line card: active, sleep, and off. Similar to the server power states,  hierarchical power states are modeled.  


Our simulator provides the interfaces to model switch-based, server-based, and hybrid network architectures. Switch-based architectures use only switches for packet forwarding; a server-only network leverages servers for both computation and switching; Finally, hybrid architectures use a combination of switches and servers for communication. Our simulator offers network configuration corresponding to several state-of-the-art topologies, including \emph{fat tree}~\cite{al2008scalable} and \emph{flattened butterfly}~\cite{kim2007flattened} for switch-based architectures, the CamCube~\cite{Abu-Libdeh:2010:SRF:1851182.1851191} topology for server-based architectures and the BCube~\cite{Guo:2009:BHP:1592568.1592577} topology for hybrid architectures.


The network module currently models communication at two levels: \emph{packet-level communication} and \emph{flow-based communication}. When dependent tasks in a job need to communicate, they can either send a single flow of data or break the flow into packets to route them. Switches will forward these flows or packets along a route of linked switches and/or servers. There is a link rate capacity associated with these links. Multiple flows or packets can simultaneously travel along a link if it has not yet been saturated. The routing path between a source and destination can be either statically generated or dynamically computed based on the communication model. 




\subsection{Job and Task Modeling}
\label{sec:task-model}

In \DCSim, each job comprises of multiple tasks that may have both spatial (e.g., multi-tiered applications) and temporal dependence (e.g., input/output dependence). Servers in the simulated environment can be configured to perform different tasks. For example, a web request can be modeled as two sequential tasks, one that is serviced by the application server and another corresponding to queries sent to database servers. Such task relationships are denoted as \emph{spatial inter-dependence}. On the other hand, \emph{temporal inter-dependence} exists when a task cannot start executing until all of its parent tasks have finished their execution, and until after their results have been communicated to the server assigned to the task. A job is considered to have finished when all of its tasks finish execution. 

Formally, each job $j$ can be represented as a directed acyclic graph (DAG) $G^j(V^j, E^j)$, where $V^j$ is the set of tasks of job $j$. In DAG, if there is a link from task $i$ to task $r$, then task $i^j$ must finish and communicate its results to task $r^j$ before $r^j$ can start processing. Each task $v^j \in V^j$ has a workload requirement, namely task size or execution time requirement $w^j_v$ for the core. For each link in $E^j$, there is a data transfer size $D^j_l$ associated with it, which denotes the bandwidth requirement to transfer the result over link $l$ (from the task at the head of DAG link to the task at the tail) when assigned a network flow. 

\subsection{Workload Modeling}
\label{wd-model}
We use two types of workload arrival models: synthetic workloads based on stochastic process, and actual system trace-based workload simulation. \DCSim currently provides two stochastic workload models:


{\it Poisson-based job arrivals}: Both the job service times and job inter arrivals are modeled as an exponential distribution with a mean service time, $1/\mu$, where $\mu$ is the service rate of a server.
In a multi-core based server farm, the relation between system utilization $\rho$ and job arrival rate $\lambda$ is: $\rho=\frac{\lambda}{\mu * nServers * nCores}$, where $nServers$ is the number of servers and $nCores$ is number of cores per server. Poisson process is widely used in prior works to model data center workloads~\cite{meisner2012dreamweaver,YaoDualDelayTimer2015}.


{\it MMPP-based bursty job arrivals}: \emph{MMPP} or \emph{Markov-Modulated Poisson Process} utilizes a continuous-time Markov chain to model different stages or states of the workload. Each state $x$ corresponds to a Poisson Process with job arrival rate $\lambda_x$. By orchestrating the transitions among various states with high and low ${\lambda}s$, \emph{MMPP} is able to model workload {\it burstiness} at a finer-grain level. We use a 2-state \emph{MMPP} model, in which one state has a high job arrival rate $\lambda_h$ representing periods of bursty arrivals, and the other state has a low arrival rate ($\lambda_l$) and models non-bursty periods of operation. There are two approaches to tune the levels of burstiness -- increasing the ratio of job arrival rates between bursty and non-bursty state, $R_a = \lambda_h/\lambda_l$, or decreasing the proportion of time the process stays in bursty state. Detailed exploration of workload burstiness modeling is a rich area of study~\cite{burstiness_analysis}. 



\subsection{Scheduling}
The simulated data center has a global scheduler which receives job requests from the front end. It is responsible for constructing a set of inter-dependent tasks corresponding to the request. The global scheduler then assigns tasks to servers based on a predetermined configured scheduling policy. We have built several global scheduling policies including \emph{round-robin} and \emph{load-balancing}. Note that the scheduling policy can be easily extended to support many modern distributed computing frameworks such as Hadoop and web services~\cite{Leverichenergyefficiencyhadoop2010}.
After a task has been assigned to a server, it will be passed on to the server's local task scheduler. The local task scheduler performs task assignment based on the availability of processor cores. It can also consider the capability of the core if heterogeneous processors are modeled. 

The global scheduler can utilize a global task queue for task assignment. Specifically, before dispatching a task to servers, it will first query the servers that are configured to serve the specific type of task. If no servers are available at that time, the global scheduler will place the task in the global queue. When servers have finished the assigned task and have available cycles, it will attempt to pull a task request from the global task queue. This scheduling model is used to model applications that have centralized control. Additionally, our global scheduler can perform task dispatching without the global queue. Under this configuration, the global scheduler first dispatches tasks to the working servers based on certain policy. The assigned server will put the task in its local task queue if no processing unit is available. 

\subsection{Power Model}

\DCSim allows users to input power profiles for various system components. The simulator also implements a few configurable power state transition controllers for the respective components. Additionally, the user can also prototype their own power policies by writing control algorithms and observing individual component's state values. 

\vspace{1mm}
\noindent\textbf{Server power profile.}
Users can derive server power profiles either by performing power measurements through configuring the system in various activity states, or by using other power estimation tools specifically for this purpose. In the former case, the power measurement can be done by reading a performance counter in the processor (e.g., Intel RAPL interface~\cite{rapl}) or by using external power meters. For the latter option, the users may use power modeling tools such as CACTI~\cite{cacti-p:iccad} and McPAT~\cite{Li:2009:MIP:1669112.1669172}. 
 
\vspace{1mm}
\noindent\textbf{Network power profile.}
Similar to server power, the user can model various components in network switches such as ports and line cards. Line cards can be considered to have their own C-state and P-state to conserve power similar to servers. Lu et al.~\cite{LuPopCornsPoweroptimization2018} derived their power model by breaking down the various architectural components of a switch and using tools such as CACTI to model switch memory power. The default C-state controller controls idle to sleep state transitions using the packet queue size as a threshold. We note that \DCSim provides flexibility to the users to implement their sleep state transition policies through considering other observable behaviors such as traffic patterns.


\section{Case Studies}

In this section, we will present several data center case studies on our simulation platform to demonstrate the capability of \DCSim. In Section~\ref{sec:resource-prov}, we show a dynamic resource provisioning policy that manages server resources according to data center loads. We then study the effectiveness of leveraging {system sleep state} with simple delay timers to enhance data center energy efficiency in Section~\ref{sec:delay-timer}. Section~\ref{sec:wasp} demonstrates fine-grained energy management using both processor and system low power states. Finally, Section~\ref{sec:popcorns} shows a server-network joint energy optimization algorithm for latency-critical workloads. 


\subsection{Data Center Resource Monitoring and Provisioning}
\label{sec:resource-prov}

\begin{figure}
	\captionsetup{font=small}
	\centering\includegraphics[width=0.4\textwidth]{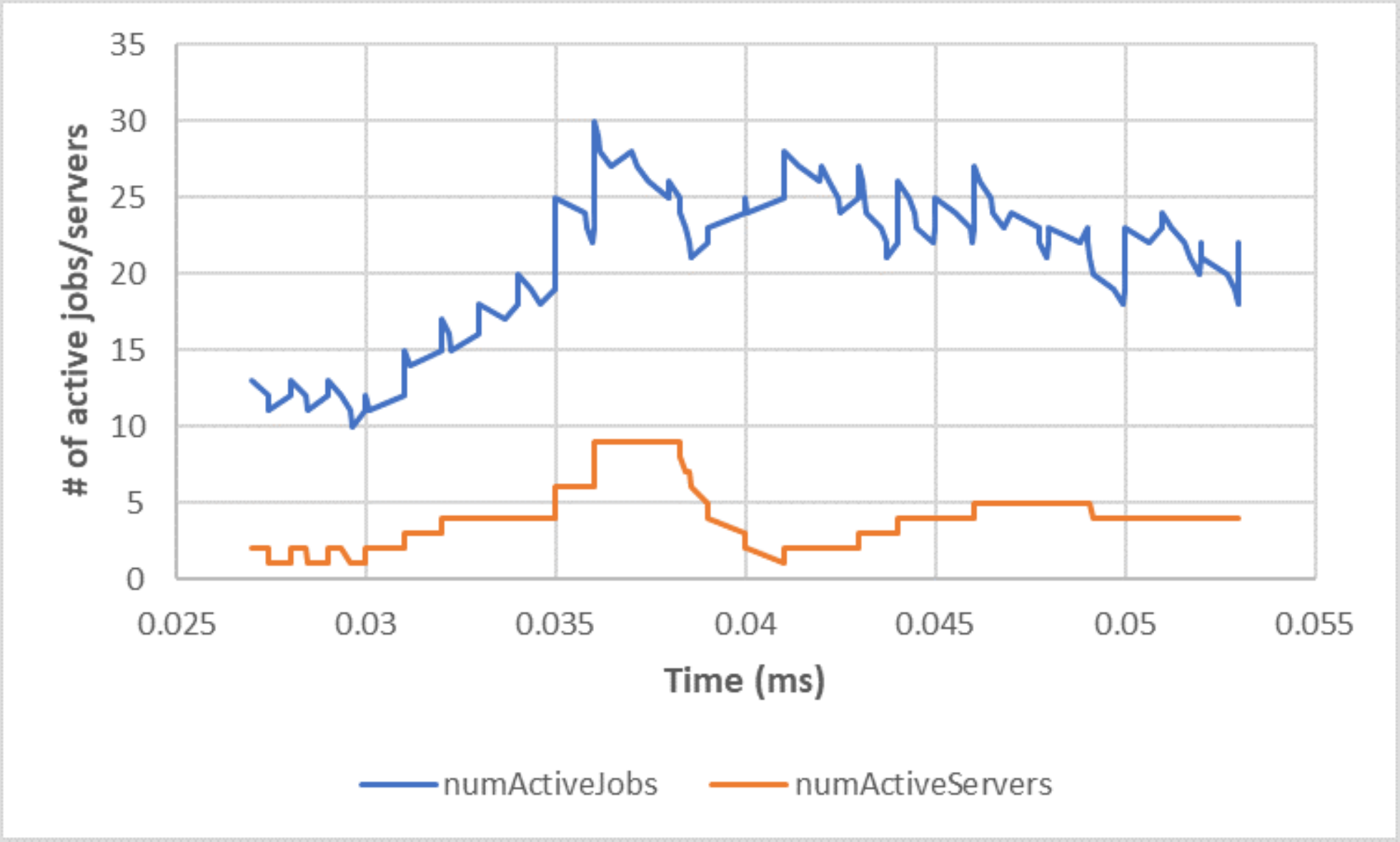}
	\caption{Number of active jobs and number of active servers over time.}
	\label{figure5}
\end{figure}

Resource provisioning is an important task for data center operators in order to budget the right amount of resources based on user demands. In this case study, we analyze the amount of servers that need to be kept active at runtime for a certain workload in \DCSim. We simulate a server farm with 50 four-core servers, and use the Wikipedia trace~\cite{wiki_trace} as our workload input. Each job consists of a simple task that has an execution time ranging from $3ms-10ms$.
Initially, all servers are in the active state. 
The global scheduler predicts the load per server in the system as it dispatches jobs to servers.
Specifically, each server is configured with a minimum and maximum load threshold. If the current load per server drops below the \emph{minimum load threshold}, one server will be put aside after finishing its pending tasks. If the current workload per server exceeds the \emph{maximum load threshold}, one server will be set to active state. 
During the initial phase, servers are gradually put to low power mode until the number of active servers stabilizes (i.e., current workload per server falls within the two thresholds). As the job arrival rate fluctuates in the system, the number of active and low-power servers will be adjusted accordingly as shown in Figure~\ref{figure5}.

Having the ability to understand active components throughout a certain period gives insight into how to properly provision resources in a data center. Using this case study, users can predetermine the number of active servers required for their performance needs to save energy costs.


\subsection{Delay Timer-based Task Scheduling}
\label{sec:delay-timer}

\begin{figure}[h]
	\centering
	\hspace{-5mm}
	\subfloat[b][Web Search]{
		\includegraphics[scale=0.30]{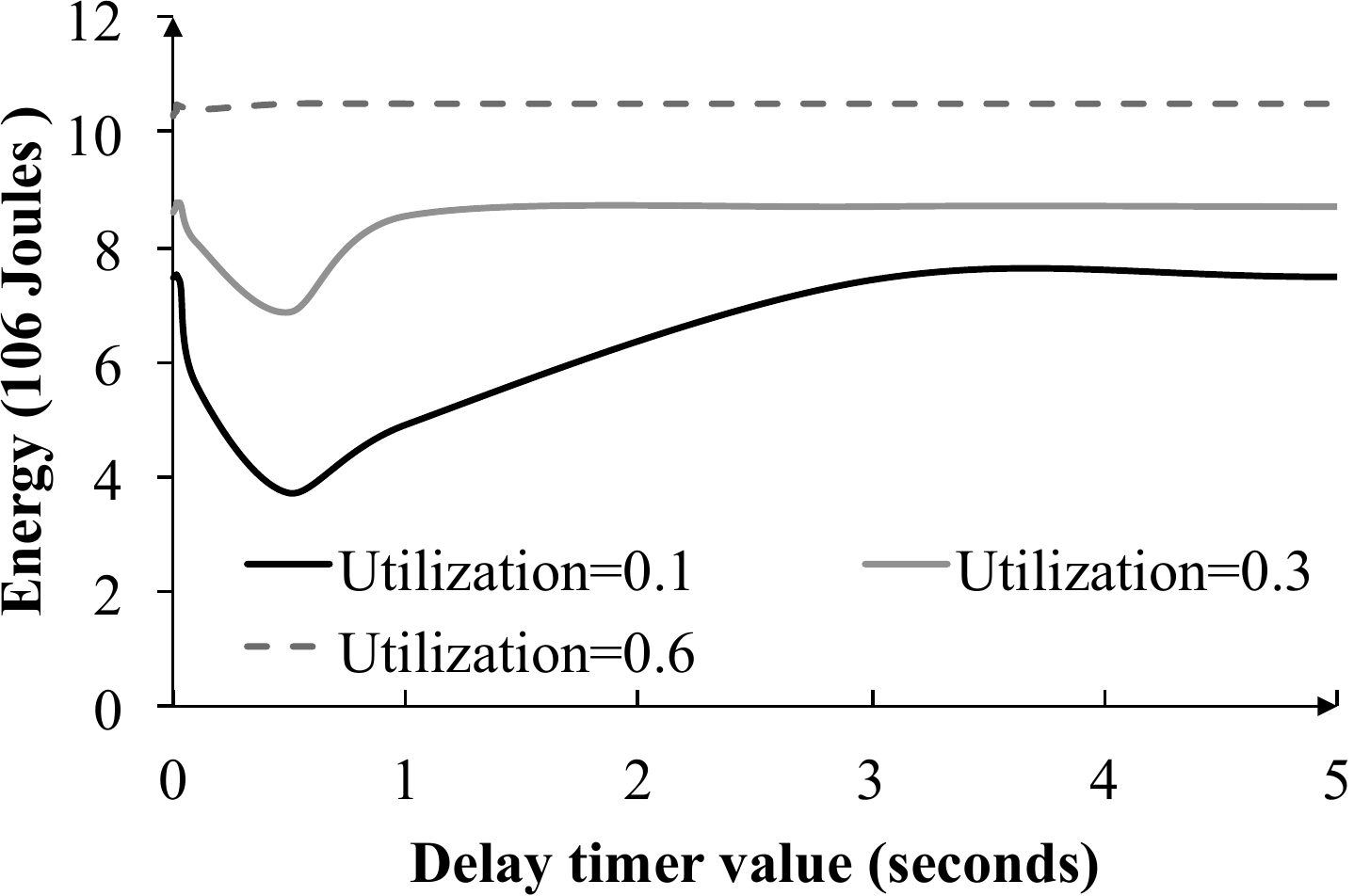}
		\label{fig:google-breakdown}}
	\subfloat[Web Serving]{
		\includegraphics[scale=0.30]{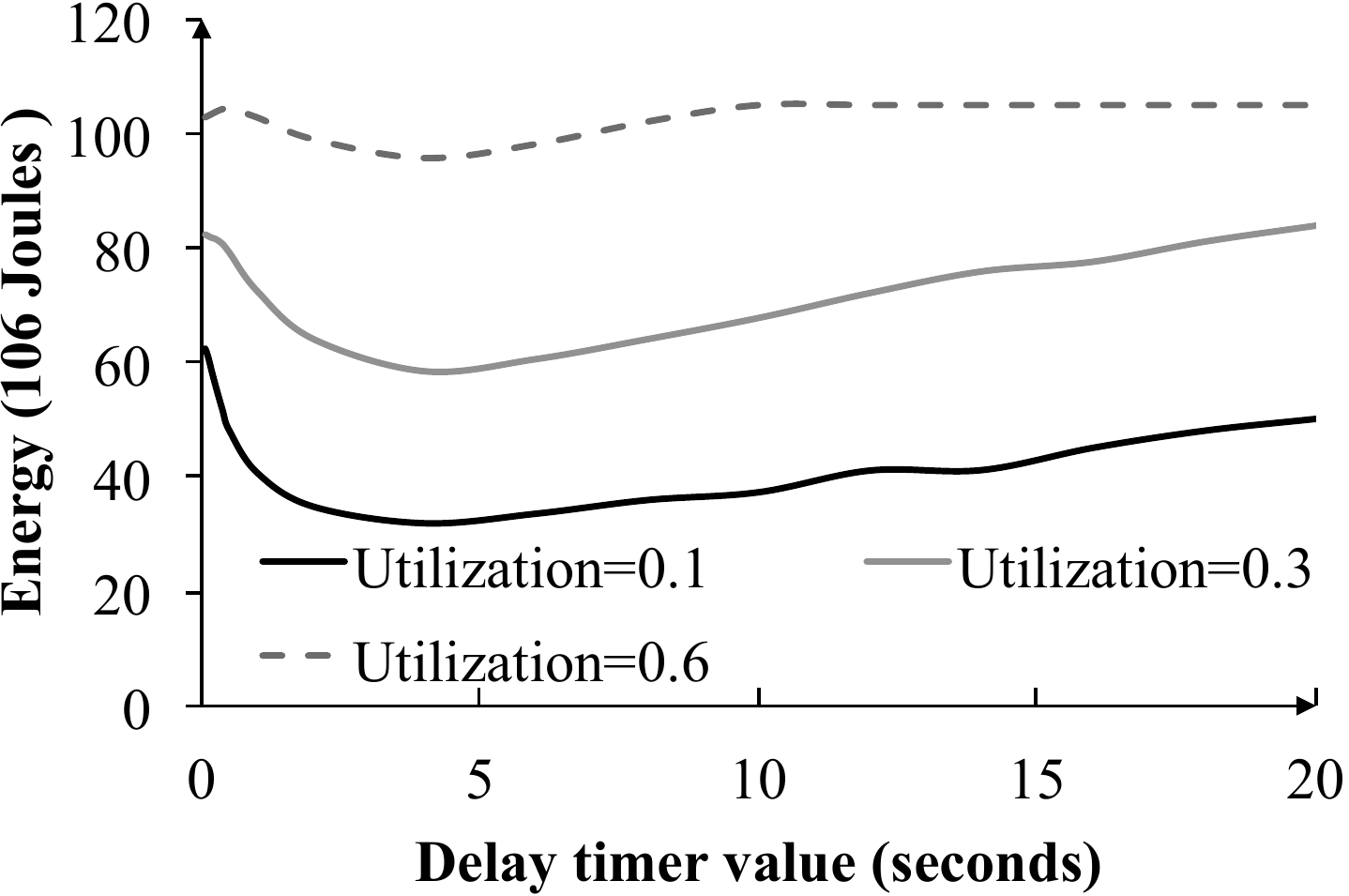}
		\label{fig:dns_breakdown}}
	
	\caption{Exploration of the single delay parameter for system on-off mechanism.}
	\label{fig:single-delay-timer}
\end{figure}
In this case study, we illustrate how \DCSim can be used as a prototype to study a server power management policy. 
One commonly studied approach is to turn off standby servers to save wasteful energy consumption. However, prior studies have shown that aggressively turning off servers can lead to even worse energy efficiency when job arrival rate fluctuates frequently~\cite{delayoff}. To tackle this issue, a possible solution is to maintain the server in highly responsible idle state for a period of time (i.e., delay timer $\uptau$) before it's turned off. However, the delay timer values need to be set judiciously in order to achieve satisfactory energy-latency trade-off.

In this experiment, we explore various settings of delay timers, and analyze their impacts on energy efficiency of two representative data center workloads. We first run simulations for servers configured with a single delay timer value.  For this simulation, we use the same server farm configuration as discussed in Section~\ref{sec:resource-prov}. \DCSim models two types of workloads: a \emph{web search} workload with relatively short service times (5ms) and a \emph{web serving} workload with longer service times (120ms). For each workload, we ran the simulation 100 times in order to explore a wide range of possible delay timer values. The exploration is performed at three different utilization levels (${\rho}$): $10\%$, $30\%$ and $60\%$. Figure~\ref{fig:single-delay-timer} shows how the settings of delay time value influence the overall energy consumption of the simulated server farm. We can see from the figure that for each workload with a fixed utilization level, there always exists one optimal delay timer setting that saves the most amount of energy. This aligns with practices that high energy consumption will occur if server are set to off state too aggressively (\emph{so that many servers end up in waking up phase most of the time}) and too conservative (\emph{which leads to many servers in standby mode even though they are not processing any tasks}). More importantly, we can observe that the optimal delay timer values are consistent across different utilizations for the same workload (0.4s for web search and 4.8s for web serving). \textbf{Our finding indicates that a single $\uptau$ value that works well across varying system loads can be set to achieve maximum energy saving,  which can significantly reduce the power management tuning efforts for system administrators.\footnote{Note that the single delay timer may not be effective when the job arrivals are highly bursty. In this case, extra server power management mechanism is needed to activate servers in time to meet application's QoS constraints.}}

\begin{figure}[t]
	\centering
	\includegraphics[scale=0.6]{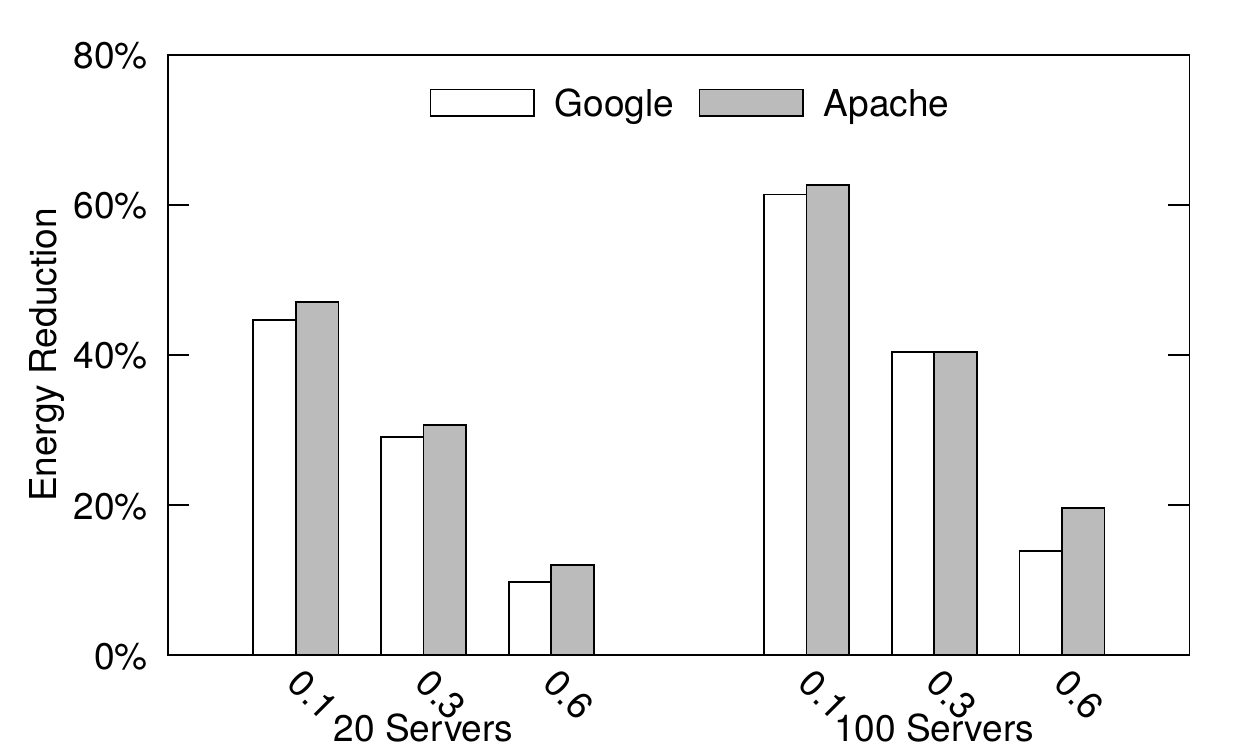}
	\caption{Energy reduction with two delay timers for web search and web serving workloads with 20 and 100 simulated servers.}
	\label{fig:energysavingsfornoofjobs}
\end{figure}

We further study the use of multiple delay timers to achieve additional energy saving, which has been proposed initially in~\cite{yao2015dual}.
The motivation is that, instead of a single $\uptau$ timer controlling transition of system low power state for each server, we could manage the servers into pools with low and high $\uptau$ times. Under this scheme, a small portion of servers with high $\uptau$ values are prioritized to process incoming jobs and kept active, while the other servers with low $\uptau$ values can quickly go to power-saving state once they finish their tasks to further save energy. Similar to the exploration of single delay timers, we run simulations on \DCSim to explore various settings including high $\uptau$ and low $\uptau$ values, and number of servers associated each of the delay timers. Figure~\ref{fig:energysavingsfornoofjobs} shows the energy saving by our proposed policy compared to a baseline policy where server are put to idle state when no tasks are assigned (Active-Idle). Our multiple delay timer mechanism can achieve upto 45\% energy reduction as compared to the baseline mechanism. Additionally, as compared to the single delay timer approach, the policy we studied can save upto 21\% energy saving while maintaining comparable job tail latencies (i.e., QoS constraints). We also observe similar energy savings when the size of server farm changes from 20 to 100. These simulations are done with minimal efforts due to the extensible implementation of our simulator which allows flexible configuration of server power management and global scheduling policies.

\subsection{Energy-Latency Optimization using Processor/System Sleep States}
\label{sec:wasp}
Modern processors incorporate a number of low power states in addition to system sleep states. These low power modes enable fine-grained power management for processors besides the system-level low power states as discussed in Section~\ref{sec:delay-timer}. 
While prior works~\cite{LiuSleepScaleRuntimejoint2014,lo2014towards} have shown the effectiveness of using processor low power states to improve server energy efficiency, they have not studied the potential of leveraging hierarchical low power states in multi-core processors (i.e., core- and package-level low power states). 
In this study, we aim to analyze the effectiveness of an energy-latency optimization mechanism that leverage both processor and system low power states to perform workload adaptive data center power management~\cite{yao2017wasp}. 

Our proposed framework performs server power management dynamically based on system-utilization at run-time. We model a load estimator that monitors the utilization of the system based on the number of pending jobs per server. Servers are coordinated between two pools. In the first group (\emph{active server pool}), servers are configured with a local power controller that only allows shallow sleep state (package C6). In the second group (\emph{sleep server pool}), each server's power controller transitions the server between shallow sleep (\textbf{package C6}) and deep sleep (\textbf{Suspend to RAM}). 

\begin{figure}[t]
	\centering
	\subfloat[b][Energy-latency optimization framework]{
		\includegraphics[scale=0.3]{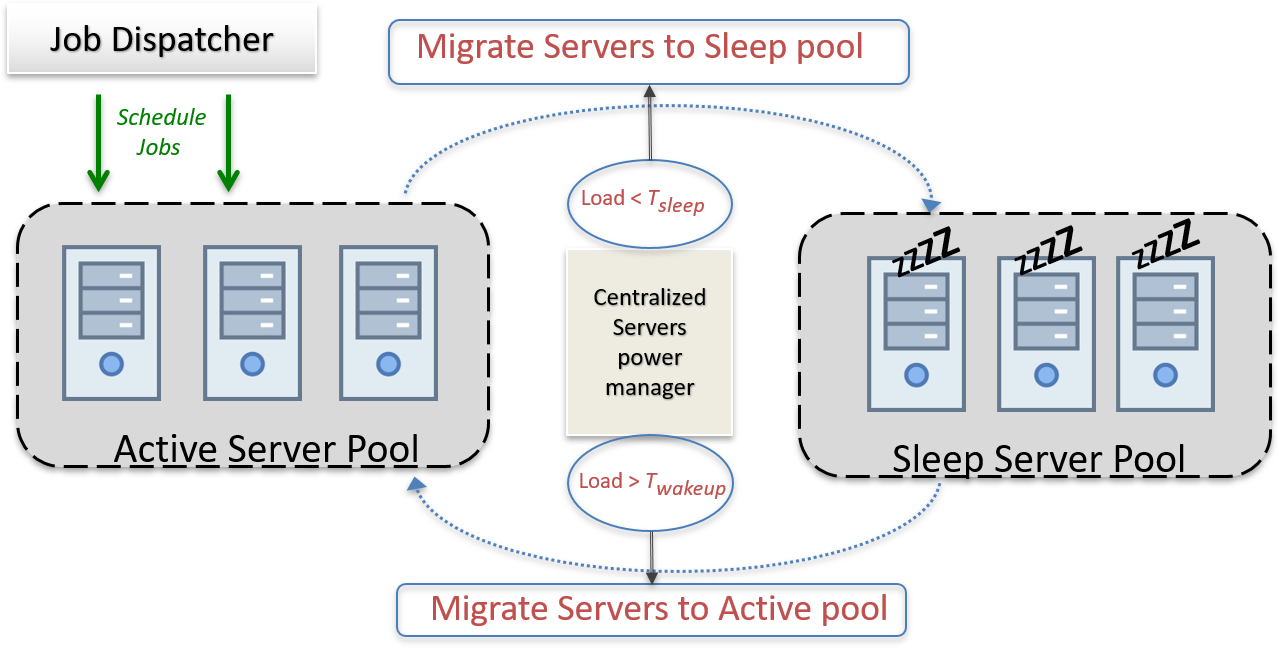}
		\label{fig:waspdesign}
	}
	
	\vspace{-4mm}
	\subfloat[b][Processor/server sleep state transitions]{
		\includegraphics[scale=0.3]{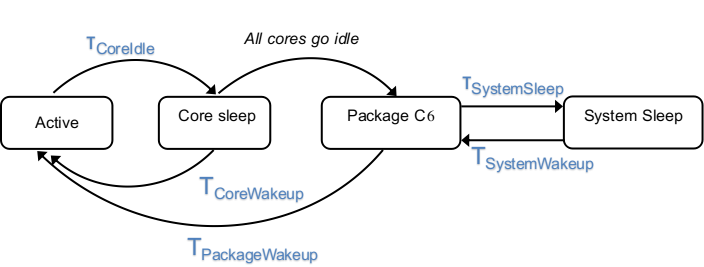}
		\label{fig:wasp_states}
	}
	\caption{The energy-latency optimization framework (top figure) and the processor/system sleep state transitions controlled by local power controller (bottom figure).}
	\label{fig:wasp_framework}
\end{figure}

To enhance energy savings, we implement a workload adaptive scheduling algorithm that dynamically transitions servers between these two pools based on current load and the target tail latency of the application. An overview of the framework is illustrated in Figure~\ref{fig:waspdesign}. Figure~\ref{fig:wasp_states} shows the core, processor and system power states transitions for these two strategies. Note that we have carefully selected core/package C6 sleep as it can bring significant power savings without long wakeup latencies (less than 1ms).
Specifically, if this load rises above the wakeup threshold ($T_{wakeup}$), a server is moved from the \emph{sleep server pool} to the \emph{active server pool}. On the contrary, when the measured metric falls below the sleep threshold ($T_{Sleep}$), the global scheduler will migrate one server from the \emph{active server pool} to the \emph{sleep server pool}. When there are bursty job arrivals, the scheduler can promptly adjust the resources in these two pool to server the requests while still saving energy for the entire data center. The front-end load balancer dispatches tasks to the servers in \emph{active server pool} only. The state of the cores, processors and the system platform are determined by each server's power controller locally. 

\begin{figure}[t]
	\centering
	\subfloat[b][Web Search]{
		\includegraphics[scale=0.5]{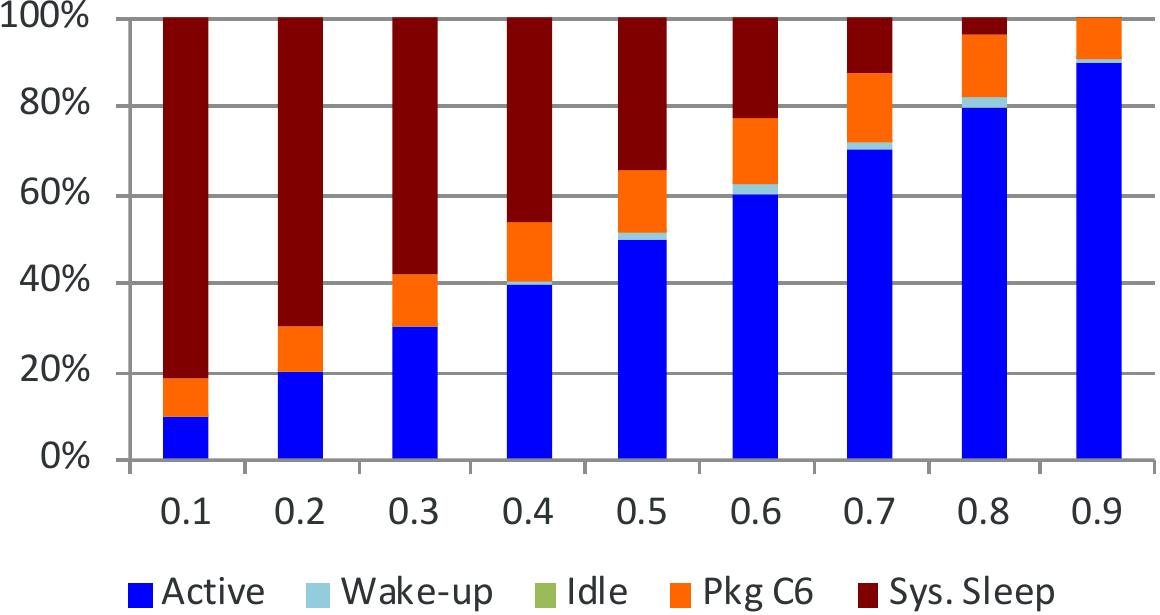}
		\label{fig:google-dist}}
	
	\vspace{-4mm}
	\subfloat[Web Serving]{
		\includegraphics[scale=0.5]{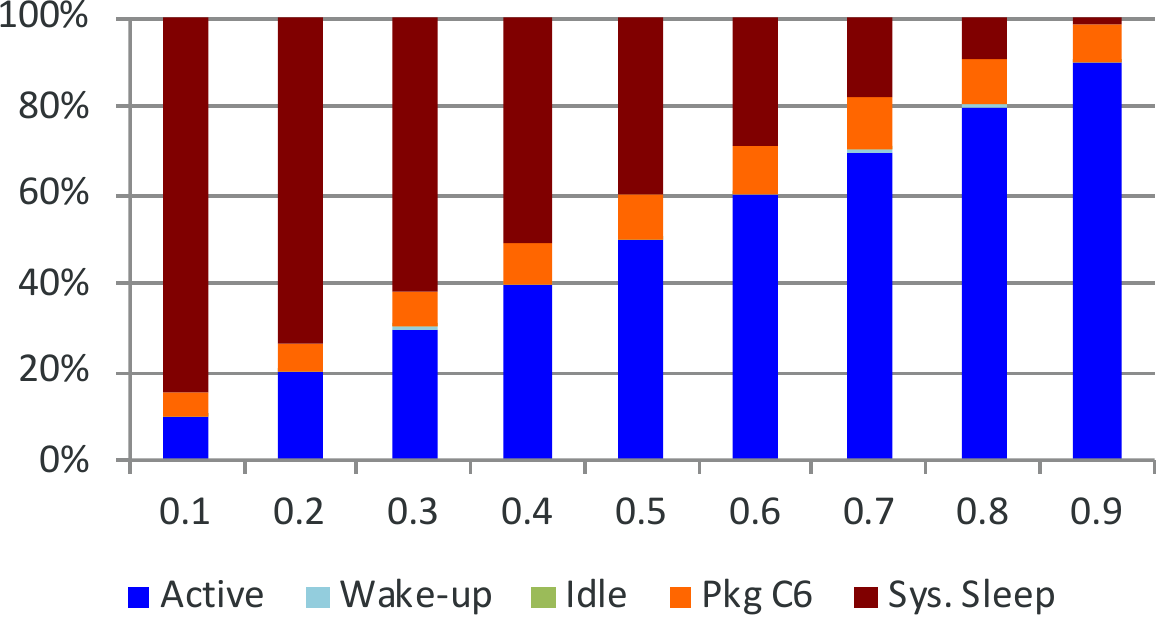}
		\label{tig:dns_dist}}
	\caption{Servers' overall state residency under the energy-latency optimization framework with different utilization.}
	\label{fig:wasp_distribution}
\end{figure}

\begin{figure}[h]
	\centering
	\includegraphics[scale=0.6]{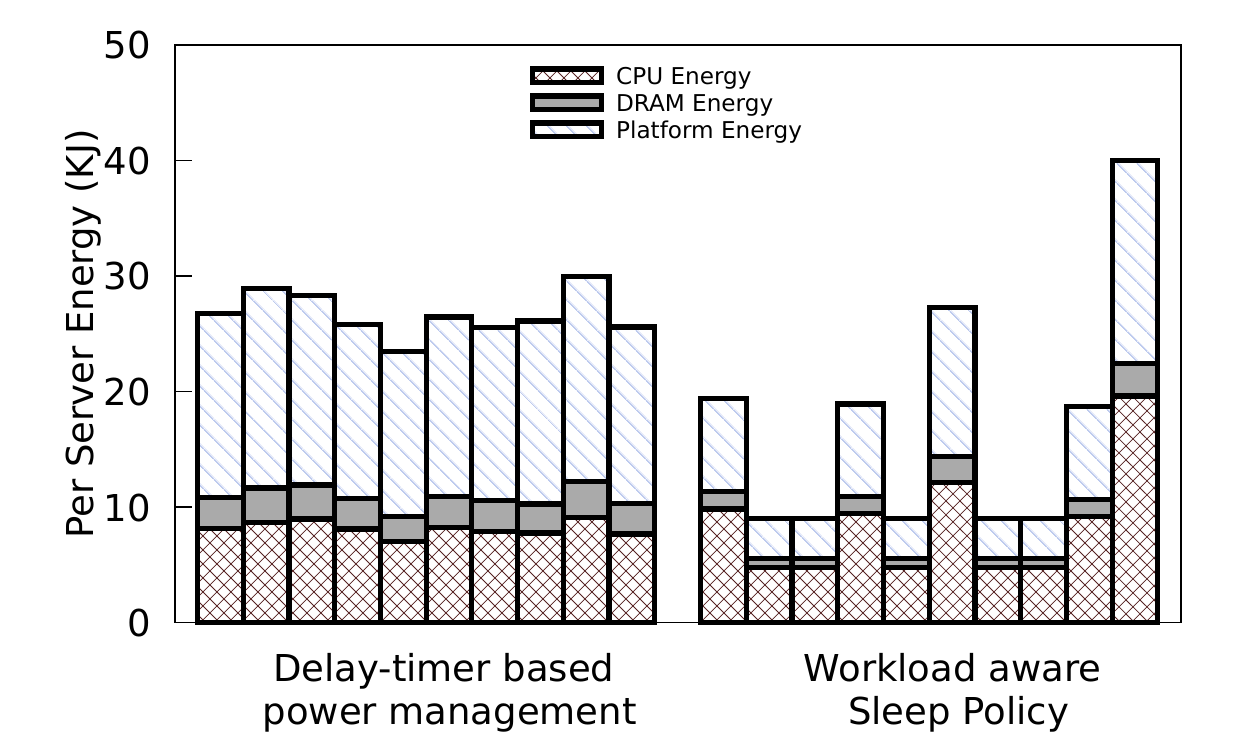}
	\caption{Energy consumption for all 10 servers using \emph{delay timer-based policy} and a \emph{workload adaptive energy-latency optimization strategy}.}
	\label{fig:waspcompare}
\end{figure}

With \DCSim, we explored the Pareto-optimal curve to analyze the trade-off between energy and achieved job tail latency ($90^{th}$ percentile) using different $T_{wakeup}$, $T_{sleep}$ and $\uptau$ values. We simulate a 10-core 10 server server farm. The processor's power and performance characteristics are based on the Intel Xeon E5-2680 processor.
We use publicly available job arrival traces, such as from Wikipedia~\cite{wiki_trace} to simulate real-world application arrival patterns. The $95^{th}$ percentile latency (QoS) is set to $2x$ the average service time for each workload.
Figure~\ref{fig:wasp_distribution} illustrates the state residency using our proposed energy-latency optimization framework for web search and web serving workloads. We can clearly see that our framework can effectively coordinate minimal amount of servers for processing tasks as the active state duration is almost the same as the system utilization. More importantly, when servers are not active, they spend most of the time in the least power-consuming state (i.e., system sleep) upto utilization of 60\%. These results indicate that the studied scheduling mechanism is highly energy-efficient. Figure~\ref{fig:waspcompare} presents the energy breakdown for each of the servers in the data center under the delay timer based approach and our workload adaptive scheduler. As compared to the delay-timer based approach that has almost uniform energy consumption across servers, the workload adaptive framework is able to automatically dispatch the tasks to a very small subset of servers (server \#6 and \#10), and keep the rest of the servers mostly in low-power states (package C6 state or system sleep state). Overall, we have seen that the workload adaptive approach can further achieve 39\% energy saving as compared to the delay timer approach. 
This case study demonstrates the comprehensive server power modeling in \DCSim, which enables future studies considering hierarchical server power management . 




\subsection{Sever and Network Cooperative Energy Optimization}
\label{sec:popcorns}
Apart from providing the capability to characterize various power saving schemes for servers, \DCSim can also be used to study techniques that co-optimize server and network energy. 
In data centers, if load balancing of network traffic is performed without considering servers, it may unnecessarily wake up servers that could otherwise be in low power states. Conversely, if load balancing is performed on servers without taking network activities into account, it may traverse excessive amount of network switches that lead to wasteful energy consumption. Intuitively, if the server job allocation algorithm is enhanced with awareness of network power assumption, we could further improve the energy consumption for the entire data center infrastructure. 


\begin{figure}[t]
	\centering
	\includegraphics[scale=0.12]{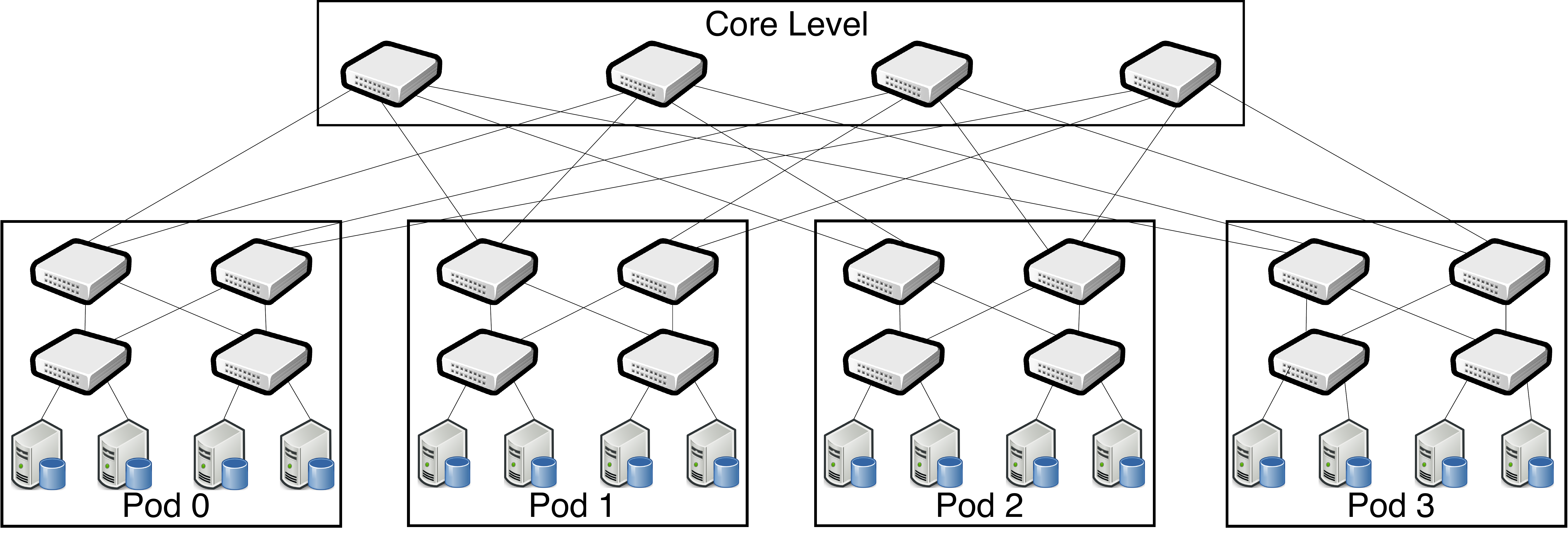}
	\caption{Network topology used in the study of combined server-network power saving strategy.} 
	\label{fig:fattree network}
	\vspace{-2mm}
\end{figure}

\begin{figure} [h]
	\centering
	\subfloat[Server and network power consumption]{
		\centering\includegraphics[width=0.8\linewidth]{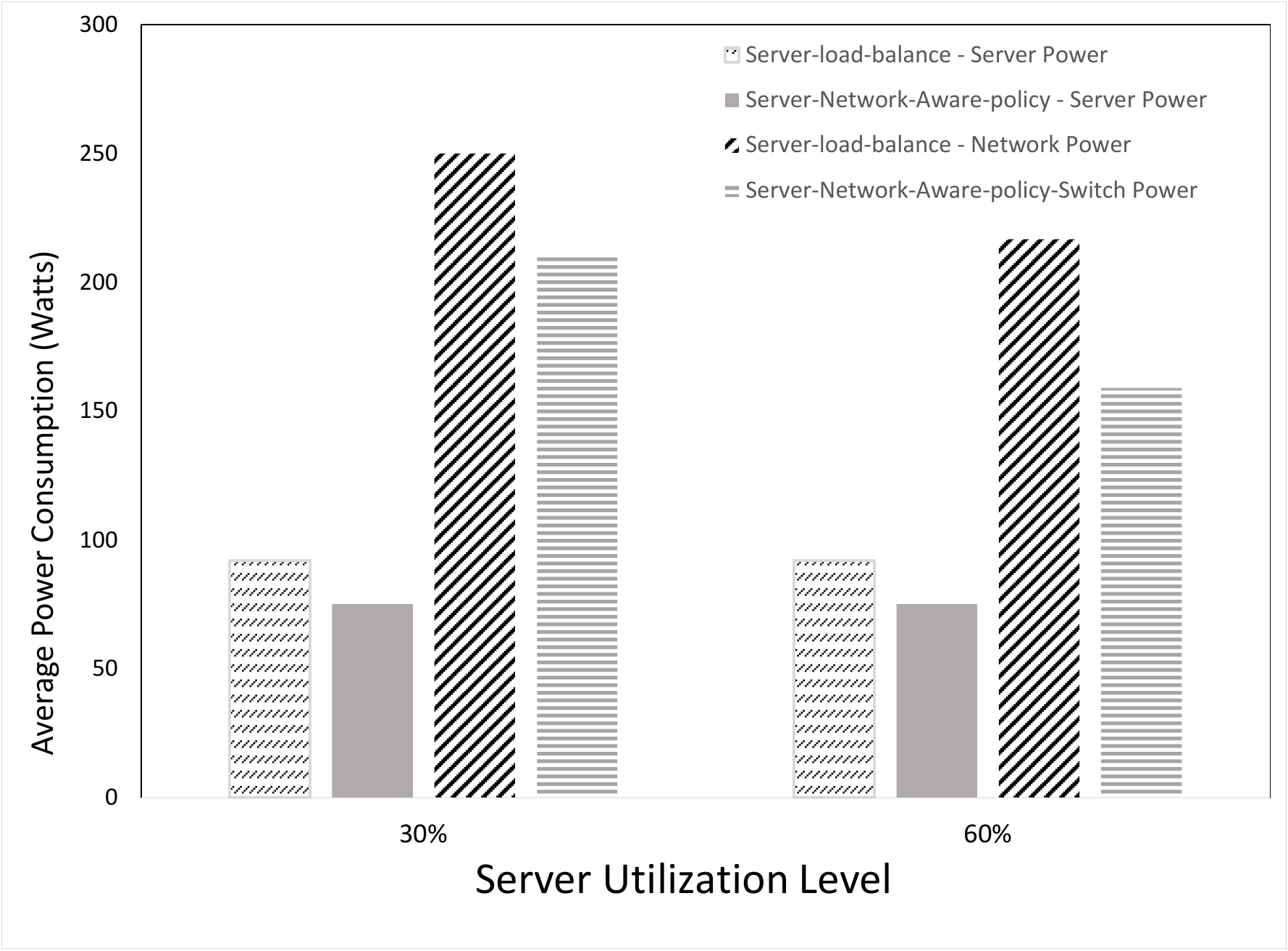}
		\label{fig:wsearchpower}
	}	
	

	\subfloat[b][Job response time]{
		\includegraphics[width=0.9\linewidth]{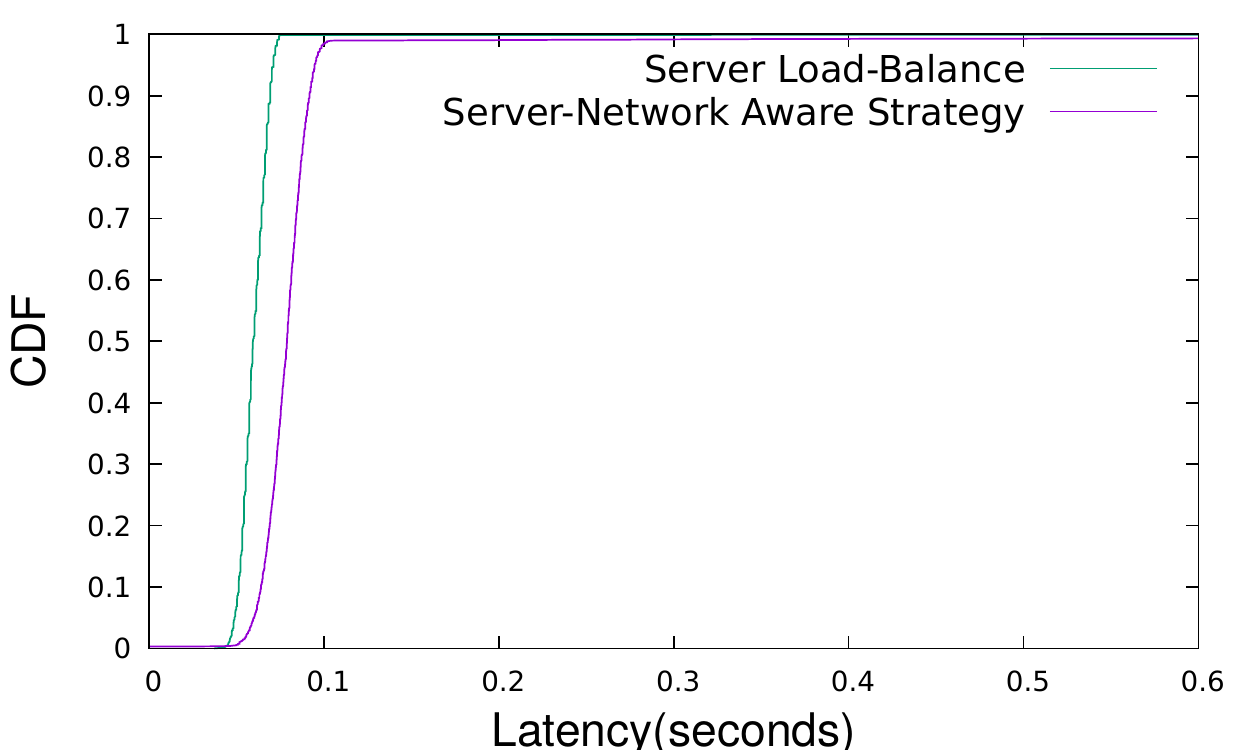}
		\label{fig:cdf_power}
	}
	
	\caption{Server and network power consumption (top figure) and job response time CDF (bottom figure).}
	\label{fig:ws-exp}
\end{figure}

In this case study, we modeled a data center using the widely deployed fat-tree topology that has full bisection bandwidth~\cite{al2008scalable}. 
To simulate network traffic, each job is simulated as a set of inter-dependent tasks as discussed in Section~\ref{sec:task-model}.
The dependence among tasks is modeled as a DAG where traffic pattern among these tasks are known. The flow size for each communication between servers is set to 100MB. We model the \textbf{Server-Network Aware} power management strategy in \DCSim that works as follows: whenever there is a need for an additional server to transit to active state, it would first identify the server with the \emph{least network cost}---the amount of additional switches to be woken up in order to allow communications to that server. For comparison, we consider \textbf{Server-Balanced} policy where jobs are strictly load balanced among servers. 
In this experiment, the network module models flow-based communication.
With \DCSim, we study the latency and power consumption for a set of jobs with randomly assigned job execution time. Figure \ref{fig:wsearchpower} presents the results for average power consumption for web search application, along with the CDF of job execution latency in figure ~\ref{fig:cdf_power} for a simulation of 2000 jobs using Poisson arrivals. We can see that we can obtain about 20\% server and 18\% network power savings with negligible increase in job latency.

\section{Validation of \DCSim Components}
In this section, we demonstrate our evaluation results of \DCSim and compare them against observations on real systems. Specifically, we show that our simulator can accurately generate the power traces for a multi-core server. We also validate the power consumption of a simulated switch via a commercial-off-the-shelf switch. 
\subsection{Server Power Validation}

In order to validate the power model of \DCSim, we set up experiments based on a 10-core Intel Xeon E5-2680 processor server. We utilize the publicly-accessible NLANR trace~\cite{nlanr}, which includes job arrival times for web service requests. We modified httperf~\cite{httperf} to enable generation of HTTP requests based on existing traces. We set up apache web service on the Xeon-based machine to serve users' requests. 
The server's power consumption is measured using Intel RAPL and IPMI interfaces~\cite{rapl,openipmi}.

\begin{figure}
	\centering
	\includegraphics[width=0.48\textwidth]{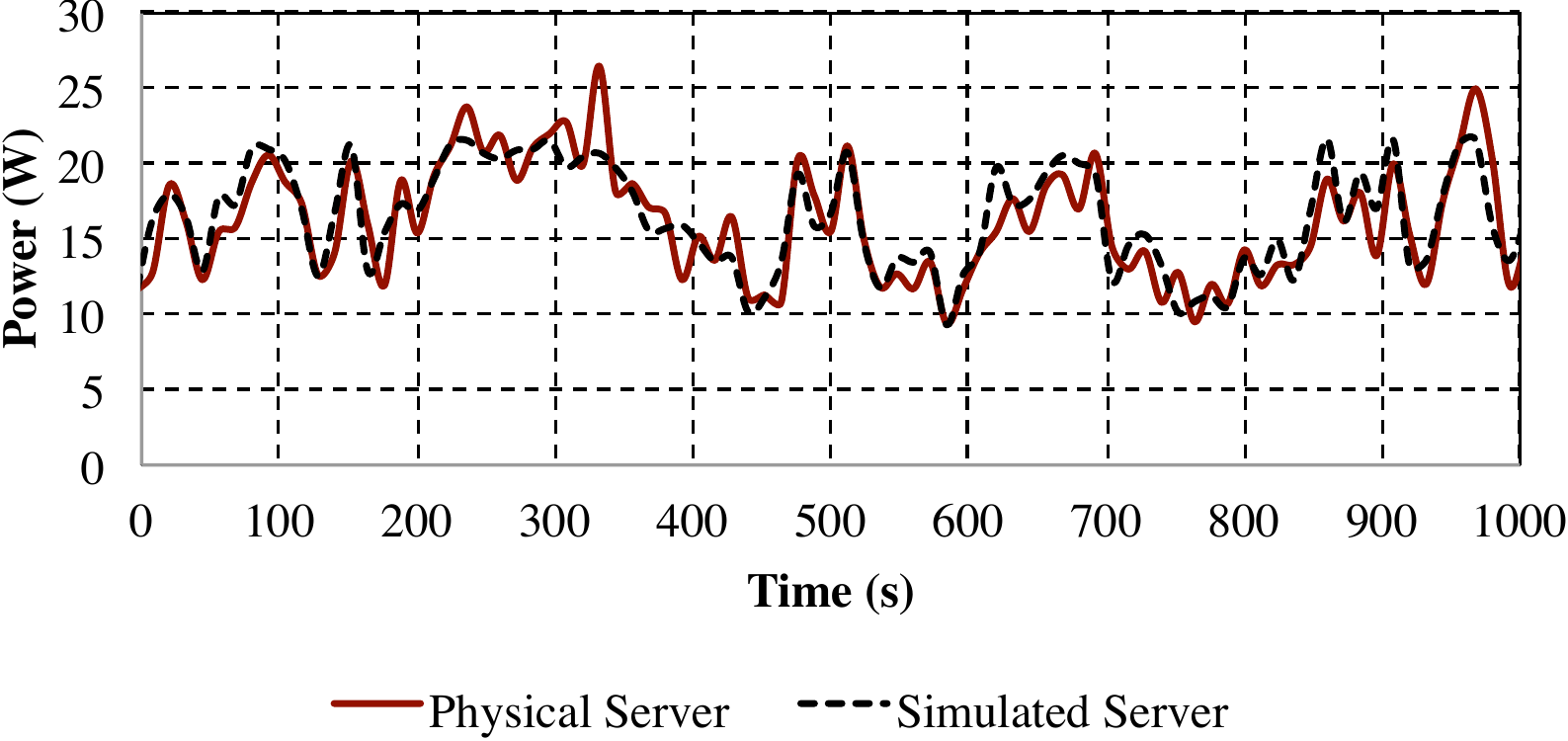}
	\caption{Power for physical and simulated server over time.}
	\label{figure17}
\end{figure}

For this experiment, we enabled two sleep states for the processor: C0 and C6. We measured the power consumption for each core as well as the entire package when they are in these sleep states. Based on the power profile, the power model for a 10-core processor server is configured in \DCSim.
We then replay the same NLANR trace in the simulator by enabling trace-based simulation. The power consumption statistics are generated and collected in \DCSim. Our validation experiment shows that the power profile generated using \DCSim matches with the power consumption curve of our physical machine with negligible error. Figure~\ref{figure17} shows a snippet of the power traces for both the physical and simulated server. We observe that the  average power difference between the physical server and the simulated server is only 0.22W, which indicates a minimal error (around 1.3\%). Additionally, the standard deviation on simulator power is only about 1.5 W. Note that on the physical server power, there are additional noise such as apache management thread and other OS routines that contribute to the difference. Overall, the patterns and values in the simulated and physical servers' power traces match closely, validating that our simulator can faithfully model servers' power consumption behavior.
\subsection{Switch Power Validation}

\begin{figure}
	\centering\includegraphics[width=0.48\textwidth]{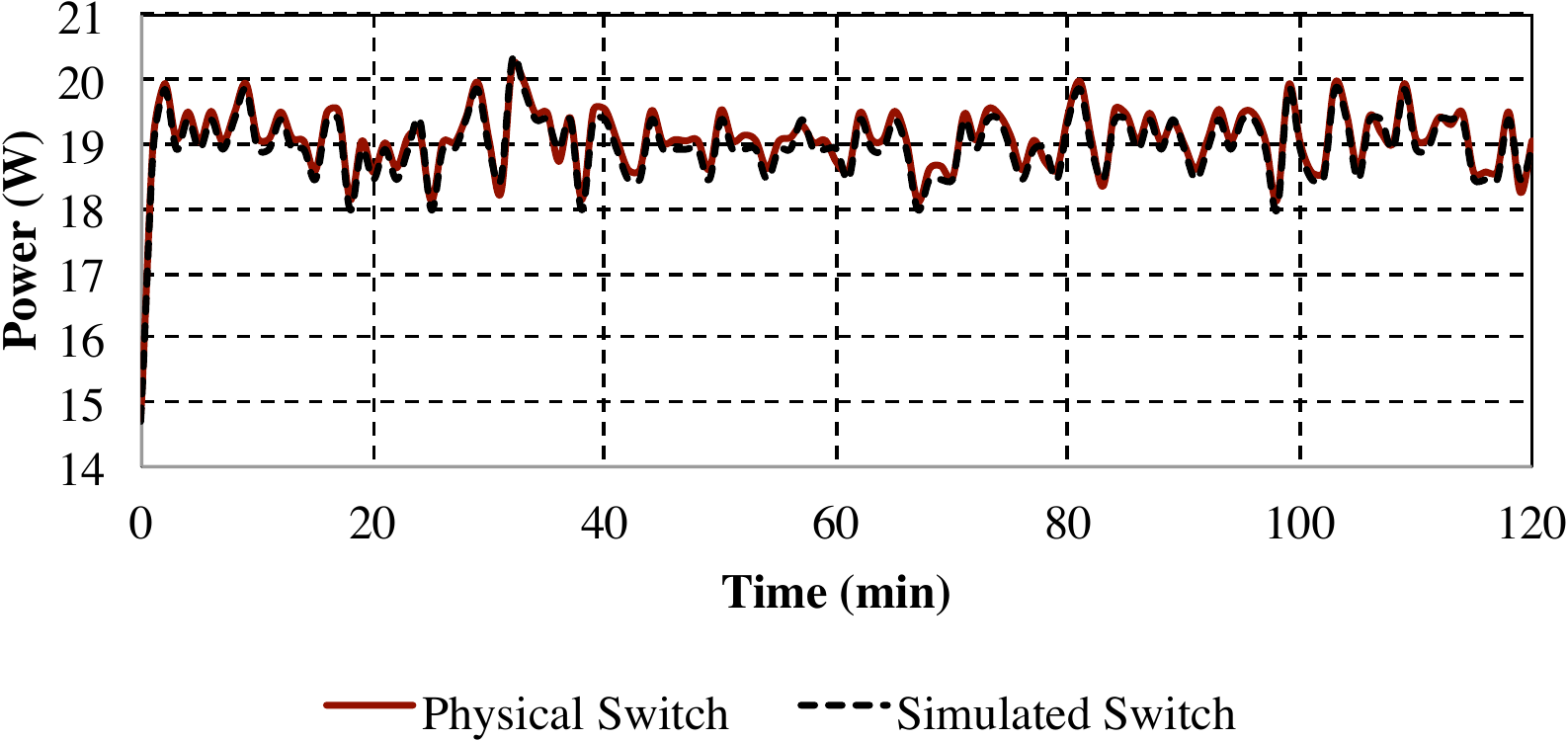}
	\caption{Power for physical and simulated switch (2-hours).}
	\label{figure14}
\end{figure}

\begin{figure} [h]
	\centering
	\subfloat[Switch power trace, Segment 1 ((80-100 mins))]{
	\centering\includegraphics[width=0.48\textwidth]{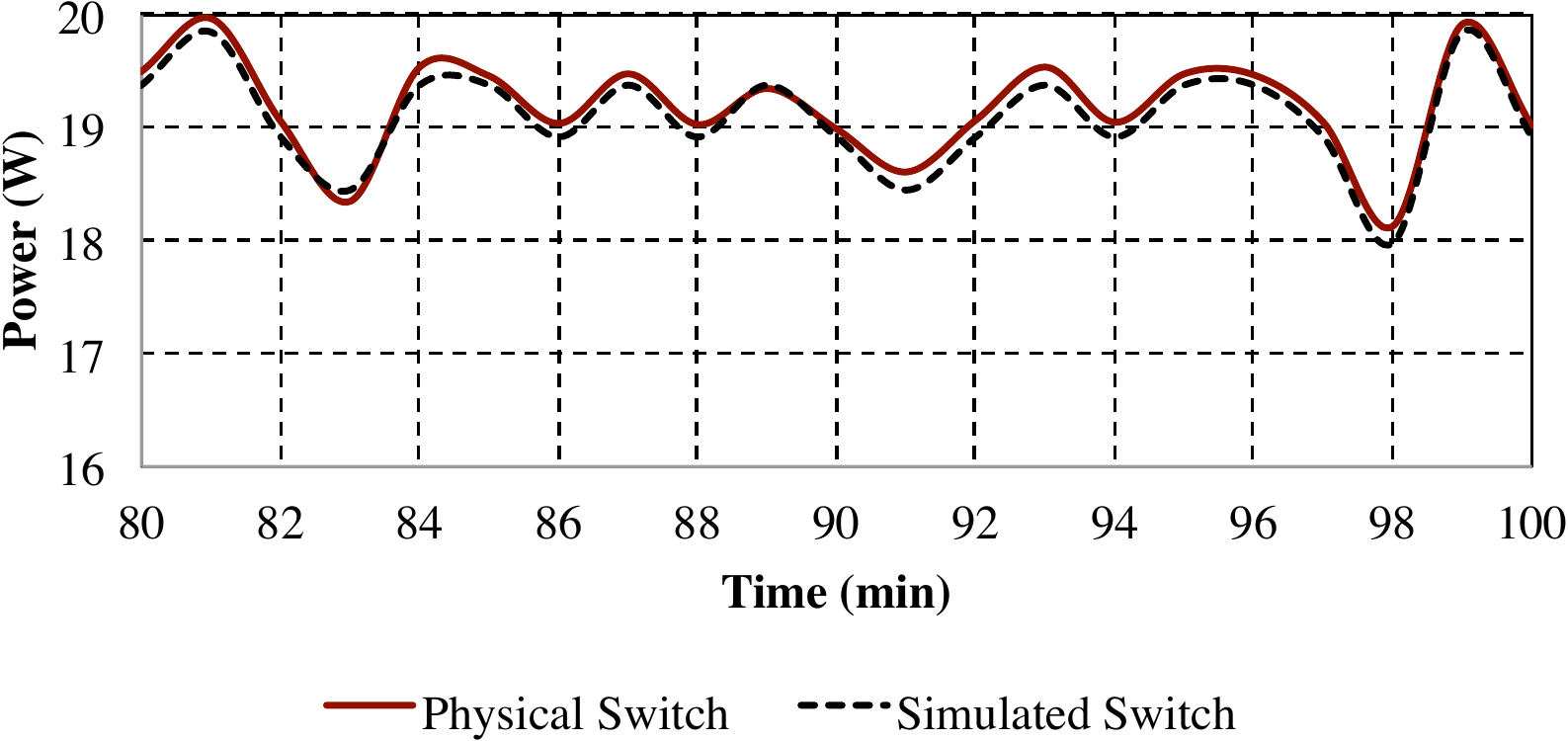}
		\label{figure15}
	}	
	
	
	\vspace{-2mm}
	\subfloat[b][Switch power trace, Segment 2 (40-60 mins)]{
		\includegraphics[width=0.48\textwidth]{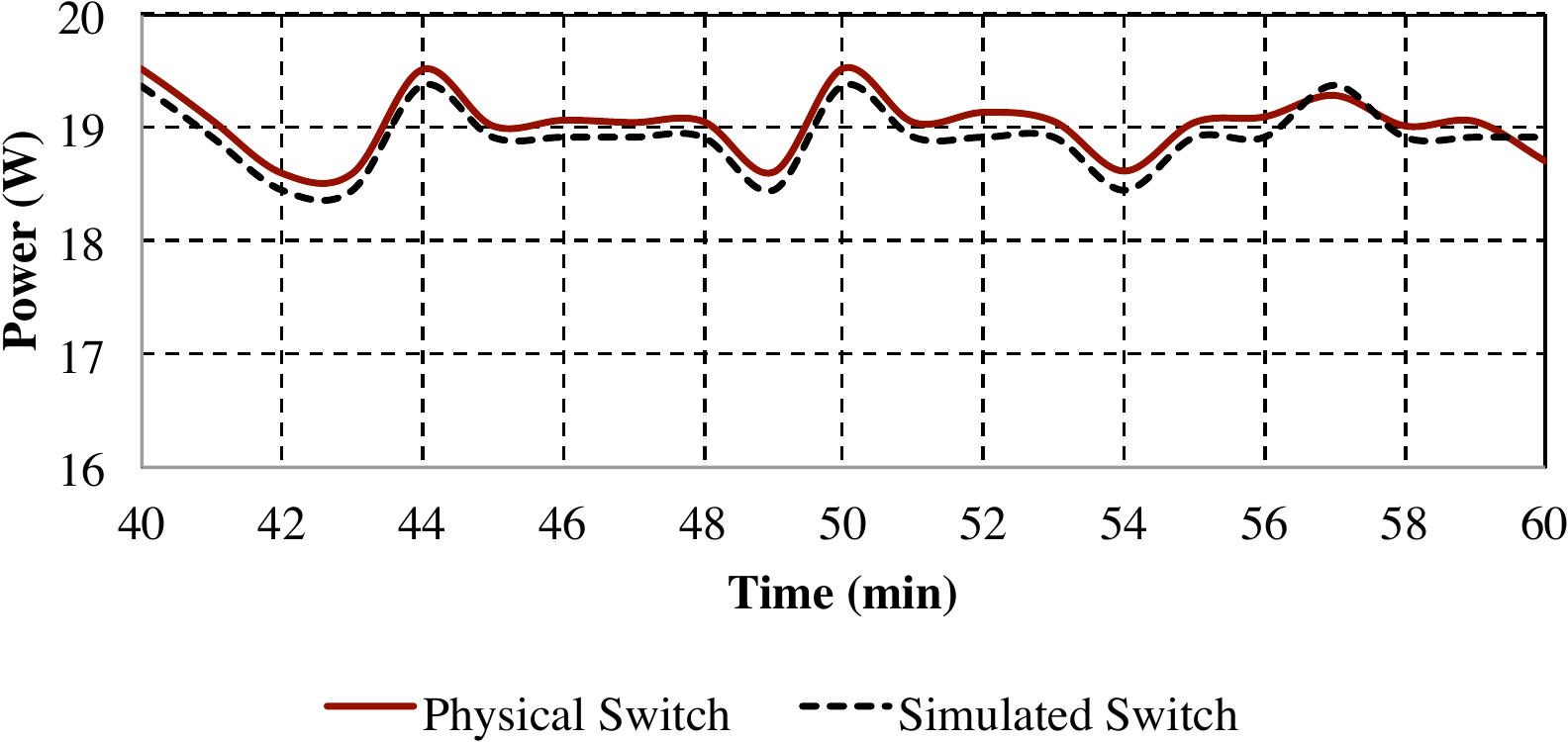}
		\label{figure16}
	}
	
	\caption{Representative segments of power traces for physical and simulated switch.}
	\label{fig:figure-segements}
	\vspace{-2mm}
\end{figure}

%

Power validation was performed on a \texttt{Cisco WS-C2960-24-S} network switch. In the simulator, we set up 24 servers connected to one switch using the star topology. We configure the switch power model using the power profile of the physical switch. 
The simulated switch has 24 ports, a base power of 14.7W, and a per port power of 0.23W. 
The cluster is configured to simulate a Wikipedia web service~\cite{wikibench} using load balanced scheduling policy. Our load generator generates user requests to the simulated cluster using the Wikipedia trace~\cite{wiki_trace}. \DCSim created a log of the port states for all the 24 ports along with the power consumption for a 2-hour simulation. 

We then implement a script for the physical switch which controls the status of the switch (including the line card and each of the ports) based on the aforementioned simulation log. We use a Power Data Logger~\cite{power-logger} to monitor the switch power by connecting the power cable directly to the Logger. 
The power consumption of the physical switch is sampled every 1 second. Figure~\ref{figure14} shows the power consumption of the simulated switch and the real switch over the 2-hour period. As we can see, the two power curves are closely tracking each other. 
We observe that the average power difference is less than 0.12W with a standard deviation of 0.04W.
In some portions of the trace, the power of the simulated server is seen to exactly match the trace of the physical switch (Figure~\ref{figure15}). In other segments of the trace, we find that the power of the physical switch is slightly higher than the simulated switch consistently, as shown in Figure~\ref{figure16}. 
In summary, we can see that \DCSim can capture the power profiles for switches fairly accurately under realistic data center workloads. 


\section{Related Work}
 
Over the past decade, many simulators are developed specifically for the prevelant computing paradigms~\cite{HowellSimJava2000,BuyyaGridsimtoolkitmodeling2002a,CalheirosCloudSimNovelFramework2009,lim2009mdcsim}. 
BigHouse~\cite{bighouse} uses a stochastic queuing model to simulate a cluster of servers with multi-core processors. The simulator is designed for studying data center server resource management policies such as resource provision and power capping. However, it does not model core and processor sleep states (such as C states) that are widely supported in many commodity servers. It also lacks network modeling, making it less effective in exploring holistic data center management. 
CloudSim is a cloud system simulator that is specifically developed to model management of virtualized resources (VMs) on physical hosts. It is also extended to support network traffic simulation~\cite{GargNetworkCloudSimModellingParallel2011}. However, CloudSim-based simulators do not consider fine-grained hardware power management features in servers and switches.
Network simulators (e.g.,~\cite{ns2,KliazovichGreenCloudpacketlevelsimulator2012}) are developed to model detailed protocols. They are typically not suitable for simulating data center scale application spanning tens of thousands of jobs.
Finally, architecture simulators~\cite{Mohammaddistgem5Distributedsimulation2017, ArgolloCOTSonInfrastructureFull2009a,SanchezZSimFastAccurate2013} allow fine-grained cycle accurate simulation by modeling hardware/microarchitecture components in detail. While these simulators are widely used to study innovations in computer architecture, they suffer scalability issues when simulating large-scale distributed systems.
Differently, \DCSim allows users to have a more holistic view of data centers by modeling both data center server and networks resources with sufficient modeling of low-level hardware characteristics, which enables fast and sufficiently accurate simulations in terms of performance, power and energy management techniques needed for future research in data centers. 


Prior works have demonstrated techniques to improve energy efficiency of servers using various power controlling knobs. 
NCAP~\cite{AlianNCAPNetworkDrivenPacket2017} proposed packet aware dynamic frequency scaling technique that save energy for latency critical workloads with bursty traffic.  SleepScale~\cite{liu2014sleepscale} studies server processor power management by orchestrating processor sleep state and frequency settings based on application's QoS requirements. Similarly, architectural studies on the switch are explored to improve performance and energy efficiency of data center network~\cite{NedevschiReducingNetworkEnergy2008, SongTrafficAwareEnergy2016,MandviwallaEnergyefficientschememultiprocessorbased2006, KannanCompactTCAMFlow2013,2011VishwanathAdaptingRouterBuffers,YaoWASPWorkloadAdaptive2017a,yao2015dual,watts-inside, yaoTSBatLeveragingTemporalSpatial,yaoTSBatProImprovingEnergy2018,lime_icse11,chen2012need,chenEnDebugHardwareSoftware2016}. 
Note that our \DCSim is designed as a lightweight, extensible data center simulator that can be leveraged to explore all these strategies. 

Finally, several recent studies considered combined network and server based job scheduling to improve energy efficiency for data centers.
Zheng et al.~\cite{2014ZhengJointpoweroptimization} propose techniques that consolidate server and network flows to reduce data center power consumption. Zhou et al.~\cite{zhouJointServerNetwork2018} demonstrate a joint server-network energy saving scheme for latency critical workload by trading off latency slack in network communication to offer more latency rooms for server side processing. By using \DCSim, these studies can be scaled to very large configurations with minimal amount of efforts.



\section{Conclusion and Future Work}

In this work, we demonstrated \emph{\DCSim}, a light-weight, holistic, extensible event-driven data center simulation platform that effectively models both server and network architectures. \DCSim is able to guide users for a variety of data center system studies including understanding workload patterns in job execution, data center resource provisioning, global and local power management, as well as detailed combined network and server performance analysis. We have demonstrated the design of our simulation infrastructure and have shown four case studies that illustrate the versatility of our framework for varied data center related tasks. 

\section*{Acknowledgments}
This material is based in part upon work supported by the National Science Foundation under Grant Number CNS-178133. Kathy Nguyen was supported through a REU supplement under the NSF award. All of this work was performed when Fan Yao and Bingqian Lu were PhD students at GWU.
\bibliographystyle{plain}
\bibliography{references}

\end{document}